# The orientation of CO intercalated between graphene and Ru(0001)

Tianbai Li and Jory A. Yarmoff*

*Department of Physics and Astronomy, University of California, Riverside, Riverside CA 92521*

## ABSTRACT

$^{13}$CO molecules are intercalated under a single layer graphene film on Ru(0001) and interrogated with helium low energy ion scattering. Single scattering is used to determine the mass distribution of atomic species visible to the ion beam and detector, and the scattering angle is varied to distinguish adsorbed from intercalated molecules. At room temperature, CO intercalates as molecules that sit upright with the O end on top, as on clean Ru. The intercalated CO tilts, more than it does on clean Ru, when the temperature is raised. This is presumably due to increased vibrational amplitudes combined with the confining effect of the graphene film.

___________________

*Corresponding author, E-mail: yarmoff@ucr.edu, phone: +1 (951) 827-5336

Graphene (Gr), a two-dimensional, single-layer sheet of $sp^2$ hybridized carbon atoms, has attracted wide attention owing to its exceptional properties, such as high electronic conductivity and chemical and thermal stability [1-5]. Chemical vapor deposition (CVD) on transition metal substrates produces large-scale, monocrystalline films with good quality [6-10] and CVD-grown graphene films can protect transition metal substrates from oxidation or corrosion due to their chemical inertness [4,11].

Small molecules, such as $O_2$ and CO, intercalate between Gr films and metal substrates, rather than adsorb on top of the graphene layer [12-14]. The molecules generally adsorb at the edge of the graphene film or at defects and then diffuse beneath the film to become intercalated at a sufficiently high surface temperature [15]. Thus, intercalation normally requires a warm sample, but it can occur at room temperature for some molecules, including CO [12]. The intercalates can act to increase the spacing between the Gr film and the substrate, decouple the film from the substrate, and modify the morphology of the Gr layer [12,16-18]. Previous work has detailed much about intercalation beneath Gr films but, to our knowledge, the orientation of the intercalated molecules has not been addressed.

CO intercalation between Gr and Ru(0001) is studied here with helium low energy ion scattering (LEIS) [19,20]. The methodology for investigating intercalation was developed in our prior study of $O_2$ exposure to Gr/Ru(0001), which found that oxygen intercalates and does not adsorb, that intercalated oxygen is less stable than oxygen chemisorbed on clean Ru, and that it desorbs and etches some of the graphene when annealed [21]. The present experiments use shadowing and blocking to determine the orientation of intercalated CO molecules in a manner not achievable with other techniques, and find that CO molecules are oriented vertically with the oxygen end pointing up but change their geometry when the sample is heated.

The experiments are performed in an ultra-high vacuum (UHV) chamber (base pressure = $4\times10^{-10}$ Torr) that contains an ion bombardment gun for sample cleaning, low energy electron diffraction (LEED) optics and the equipment needed for LEIS. The 1 cm diameter sample is mounted on the foot of an x-y-z manipulator that enables rotations about the polar and azimuthal angles. An e-beam heater filament is located behind the sample and the temperature is measured by type K thermocouples.

The Ru(0001) is cleaned with ion bombardment and annealing (IBA) plus an oxygen treatment [22,23]. A one-hour 500 eV $Ar^+$ ion sputtering at a flux of $4\times10^{13}$ ions $sec^{-1}$ $cm^{-2}$ is followed by annealing under $4\times10^{-8}$ Torr of $O_2$ at 1100 K for 8 min to chemically remove carbon contaminants and then a flash annealing at 1300 K under UHV for 2 min to remove the oxygen residue. This cleaning cycle is performed several times to acquire a clean and well-ordered surface, as monitored with LEIS and LEED.

The graphene layer is grown via CVD by heating the Ru(0001) sample to 900 K under $1.5\times10^{-7}$ Torr of ethylene for 5 min, flash annealing under vacuum at 1200 K for 1 min, and slow cooling down to 450 K for another 5 min [23]. The process is repeated until a fully-covered, single and continuous graphene layer is formed, as verified with LEIS and LEED.

After graphene growth, exposures of $^{13}CO$ are conducted at pressures of $8\times10^{-2}$ Torr on Gr/Ru(0001) and $1\times10^{-5}$ Torr on bare Ru(0001) with the samples at room temperature. Exposures are reported in Langmiurs (L), where 1 L = $1\times10^{-6}$ torr sec.

A differentially pumped ion gun creates a beam of 3.0 keV He$^+$ for LEIS that has a diameter of approximately 2 mm and a flux of 1.9×10$^{11}$ ions sec$^{-1}$ cm$^{-2}$. Scattered ions are collected by a hemispherical electrostatic analyzer (ESA) mounted on a rotatable platform, which enables adjustment of the scattering angle. A specular geometry is used in which the incident and exit angles are equal with respect to, but on opposite sides of, the surface normal. Each spectrum is collected for 86 s, which leads to a fluence that is equivalent to about 1.6% of a monolayer (ML). Although spectra collected successively from the same Gr-covered sample show no changes due to beam damage, the samples used here are re-prepared after collecting each spectrum to be certain [21].

LEIS data can be analyzed with the binary collision approximation (BCA), which assumes that the projectile interacts with only one target atom at a time [24]. The target atoms are considered unbound, as the projectile kinetic energy is much higher than bonding energies. The main features in LEIS spectra are single scattering peaks (SSPs), which correspond to projectiles that experience one hard collision with a target atom before escaping the surface and reaching the detector. The SSPs ride atop a small background of multiply scattered ions. The energy loss during a single collision is determined primarily by the target to projectile mass ratio and the scattering angle [19]. Thus, the kinetic energy and intensity of the SSPs provide the elemental composition of the near-surface region. For the data shown here, the SSPs for $^{12}$C, $^{13}$C, $^{16}$O and Ru have the expected kinetic energy calculated using the BCA and those energies change with scattering angle accordingly.

Helium projectiles are used to enable single scattering at a large angle from the relatively light C and O atomic species. Helium also has an extremely high surface sensitivity due to Auger neutralization (AN), which is an irreversible process that dominates charge transfer between noble gas projectiles and solid targets. Most of the projectiles that experience multiple collisions or collide with deeper lying target atoms undergo AN and are thus not detected [19,25]. There is also a matrix effect for He$^+$ scattered from graphitic carbon due to a quasi-resonant process in conjunction with AN that leads to nearly complete neutralization for ions with kinetic energies below 2.5 keV [26-28]. To avoid the matrix effect and guarantee the detection of scattered projectiles, 3.0 keV He$^+$ ions are employed here.

Helium LEIS spectra collected from as-grown Gr/Ru(0001) show only a sharp $^{12}$C SSP and no Ru SSP [21], which indicates the formation of a complete graphene overlayer. LEED patterns display satellite spots that are caused by the superlattice formed between the Gr film and the substrate [18,21].

Shadowing and blocking are unique features of LEIS [19] that can be used to distinguish intercalated atomic species from those of the Gr layer and any adsorbates [21]. Shadowing occurs when a surface atom sits above a deeper lying atom so that it cannot be directly impacted by incoming ions. A shadow cone is the region behind an impacted atom inside of which other atoms cannot be hit directly by the incident ion beam. Blocking is a similar phenomenon that prevents ions scattered from deeper lying atoms from reaching the detector because a surface atom is positioned above it. Figure 1 shows the sizes and orientations of the shadow cones formed at the two scattering angles used here. The size of the shadow cones was calculated using the formula in Ref. [20]. At the smaller angle of 45°, the incident and exit trajectories are close to the surface plane causing any atoms below the graphene overlayer to be completely shadowed so that He can only singly scatter from $^{12}C$ in the graphene film and any adsorbates attached atop of it. For the larger 115° scattering angle, the trajectories are closer to the surface normal so that the ions penetrate more deeply and can singly scatter from atoms below the Gr overlayer.

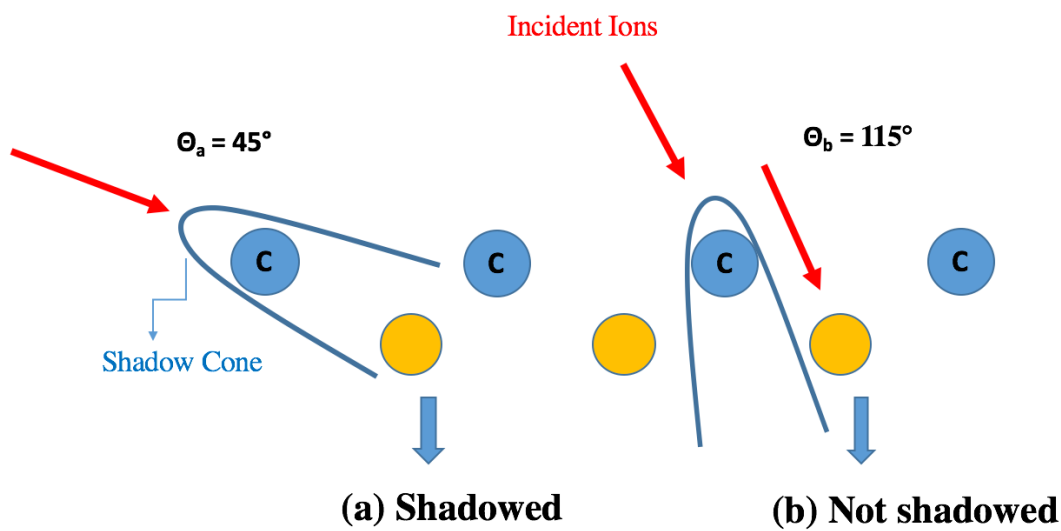

**Figure 1.** Schematic diagram illustrating shadowing during ion scattering from a graphene film at scattering angles of (a) 45° and (b) 115°. In (a), all of the atoms underneath the graphene overlayer are within the shadow cone produced by the Gr atoms and therefore cannot be directly impacted by the incident ions. In (b), atoms in the second layer are not shadowed so that the incident ions can impact intercalated molecules and the substrate.

Shadowing and blocking are used here to determine the orientation of intercalated CO molecules. To distinguish the $^{12}C$ species in graphene from carbon in intercalated molecules, isotopically enriched $^{13}CO$ is used. LEIS spectra collected from Gr/Ru(0001) exposed to $1\times10^9$ L of $^{13}CO$ are shown in Fig. 2(a), with the different scattering angles used to locate the $^{13}CO$ molecules [21]. In the spectrum collected at 45°, only the $^{12}C$ SSP from Gr is visible, which leads to the conclusion that $^{13}CO$ intercalates beneath the Gr film and does not adsorb atop the overlayer. At the 115° scattering angle, however, an additional $^{16}O$ peak is present and no $^{13}C$ is detected. This indicates that the ¹³CO is intercalated but is oriented vertically with the oxygen atoms on top such that they shadow the ¹³C.

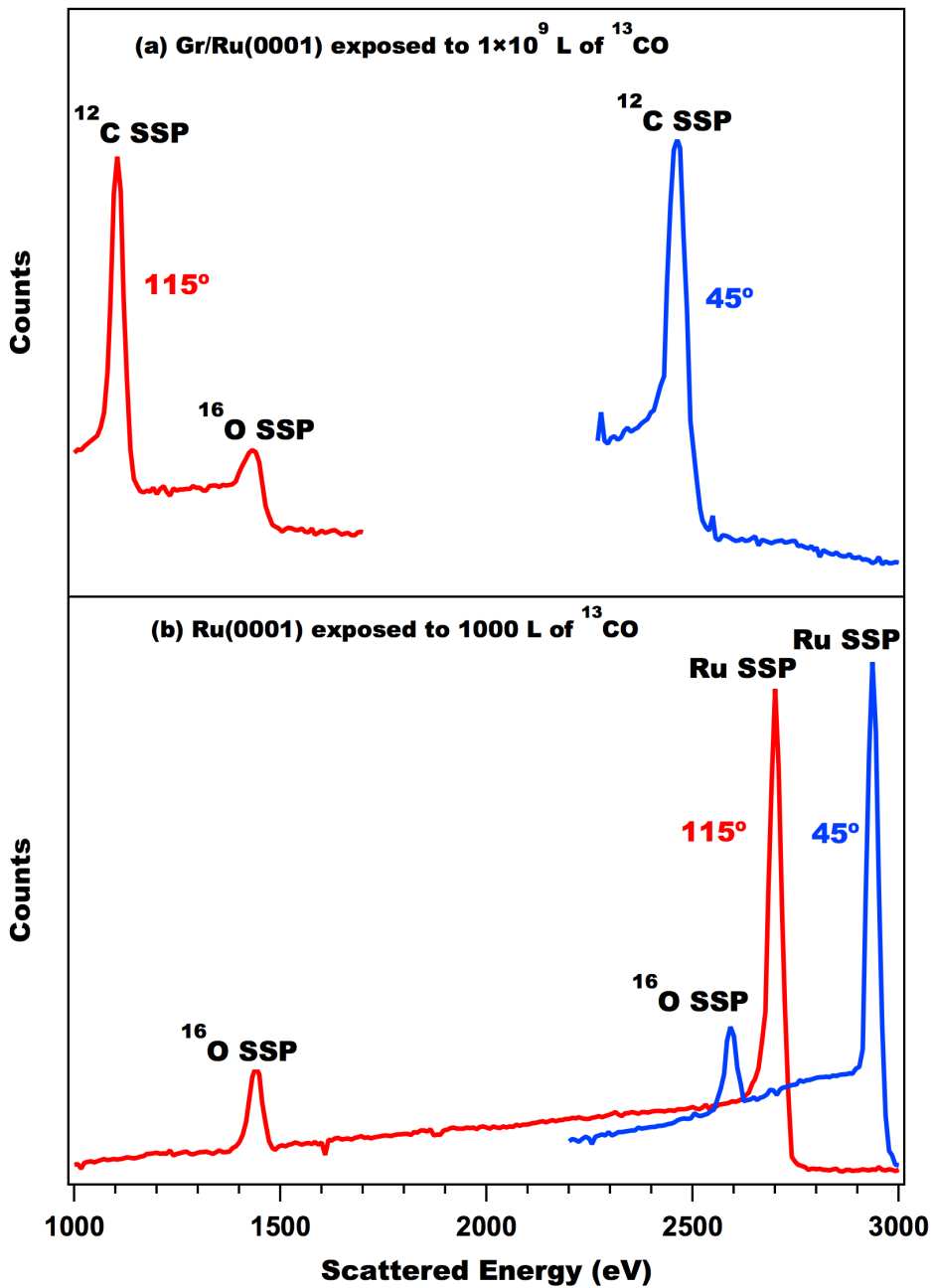

**Figure 2.** 3.0 keV He⁺ LEIS spectra collected at scattering angles of 45° and 115° from (a) Gr/Ru(0001) exposed to $1\times10^9$ L of ¹³CO and (b) bare Ru(0001) exposed to 1000 L of ¹³CO.

Spectra collected from a saturation coverage of 0.6 ML of $^{13}$CO on bare Ru(0001) are shown in Fig. 2(b). Only $^{16}$O and Ru SSPs are observed at both scattering angles. The absence of a $^{13}$C SSP at 115° indicates that $^{13}$CO adsorbed on Ru(0001) is also adsorbed vertically with the O end up, consistent with the literature [29]. A $^{13}$C SSP is not observed at 45° because the high CO coverage leads to shadowing and blocking by neighboring adatoms.

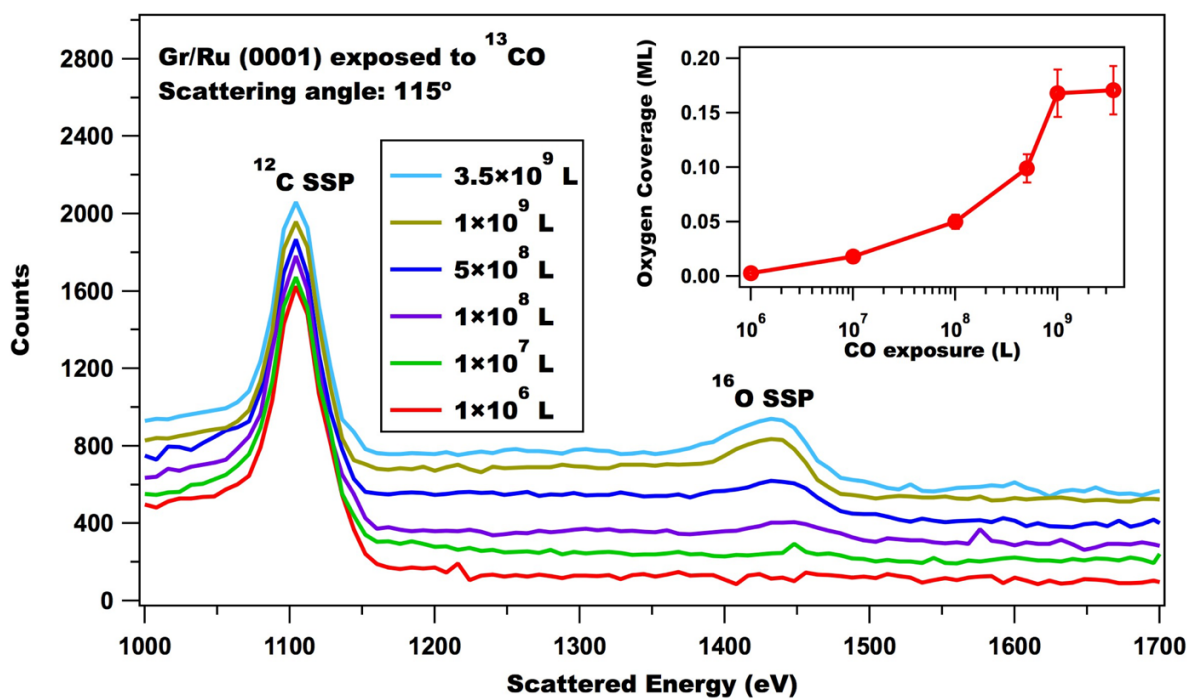

**Figure 3.** 3.0 keV He$^+$ LEIS spectra collected at a scattering angle of 115° from Gr/Ru(0001) following the indicated $^{13}$CO exposures. The inset shows the oxygen coverage as a function of $^{13}$CO exposure.

Figure 3 shows spectra collected at 115° as a function of exposure. The oxygen SSP does not appear until a rather large exposure of 1×10$^8$ L, indicating that the efficiency of CO intercalation at room temperature is rather low. The $^{13}$CO concentration increases with additional exposure until it saturates at ~1×10$^9$ L. Note that Ref. [30]

reported that CO does not intercalate beneath a complete Gr layer on Ru(0001), but their exposures were at significantly lower pressures.

The inset in Fig. 3 shows the oxygen coverage quantitatively as a function of exposure. Coverages are calculated here and below from the areas of the relevant SSP by assuming that the $^{12}$C SSP from as-grown Gr/Ru(0001) represents a full ML of graphene and that the SSP areas are proportional to coverage. The SSPs are integrated after subtracting the multiple scattering background, which is modeled via a polynomial fit of the region surrounding the peak, and the ratios then are normalized by the differential scattering cross sections [20]. The coverages might be underestimated, however, because some intercalated atoms can be shadowed by the graphene overlayer and the neutralization in scattering from buried atoms could be larger than from C in the Gr film. Nevertheless, the trends of how coverage evolves with exposure and temperature are correct.

Spectra collected from $^{13}$CO intercalated Gr/Ru(0001) as a function of surface temperature are shown in Fig. 4. The uppermost solid line (2$^{nd}$ from the top) shows the LEIS spectrum collected at room temperature from Gr/Ru(0001) exposed to $1\times10^9$ L of $^{13}$CO, which contains $^{16}$O and $^{12}$C SSPs similar to Fig. 3. The spectra shown below this one were collected with the $^{13}$CO-exposed sample held at the indicated temperature for at least 120 sec. The inset of Fig. 4 shows $^{13}$C and $^{16}$O coverages calculated from these spectra as a function of temperature. The $^{16}$O coverage displays a stepwise decrease with temperature, while $^{13}$C begins to appear at ~330 K, reaches a maximum coverage at ~365 K and then decreases at higher temperatures. The first step-like decrease in the $^{16}$O coverage corresponds with the appearance of the $^{13}$C SSP. This is interpreted as resulting from a change in the geometry of the $^{13}$CO molecules with temperature that reveals the underlying $^{13}$C atoms, meaning that the molecules are lying flat or are at least substantially tilted. The second slope change occurs at ~370 K when both SSPs begin to disappear, indicating thermal desorption of the intercalated $^{13}$CO.

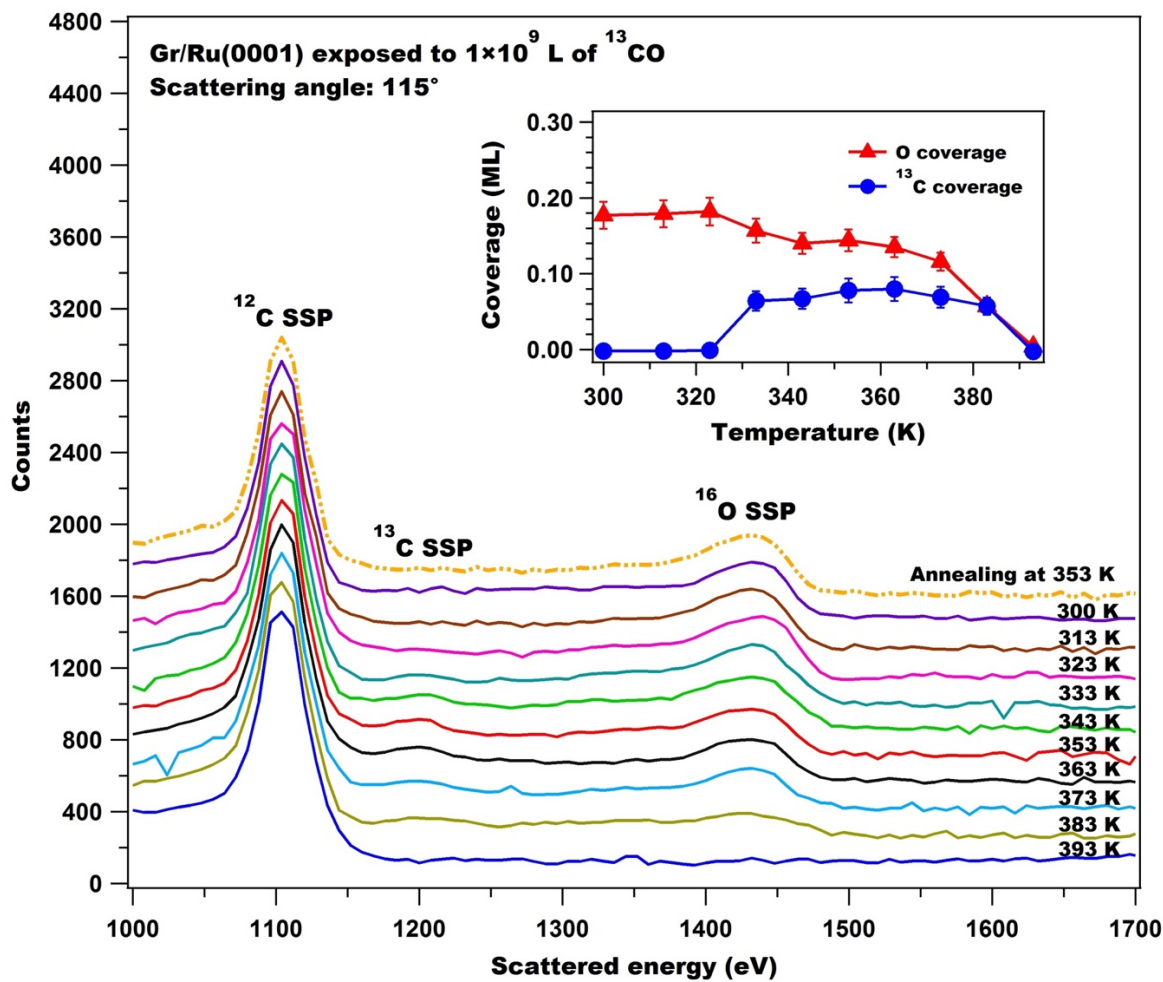

**Figure 4.** The solid lines show 3.0 keV He$^+$ LEIS spectra collected at a scattering angle of 115° from Gr/Ru(0001) exposed to 1×10$^9$ L of $^{13}$CO at room temperature and then heated to the indicated temperatures. The uppermost spectrum (dashed-dotted) was collected from Gr/$^{13}$CO/Ru(0001) after annealing at 353 K and then cooling back to 300 K. The spectra are offset from each other for clarity. The inset shows the $^{13}$C and $^{16}$O coverages as a function of the sample temperature.

In addition, $^{13}$CO-exposed Gr/Ru(0001) was annealed to 353 K and the uppermost spectrum in Fig. 4 was collected after letting the sample cool back to room temperature. This spectrum is similar to that obtained before heating, showing that the appearance of the $^{13}$C SSP is reversible.

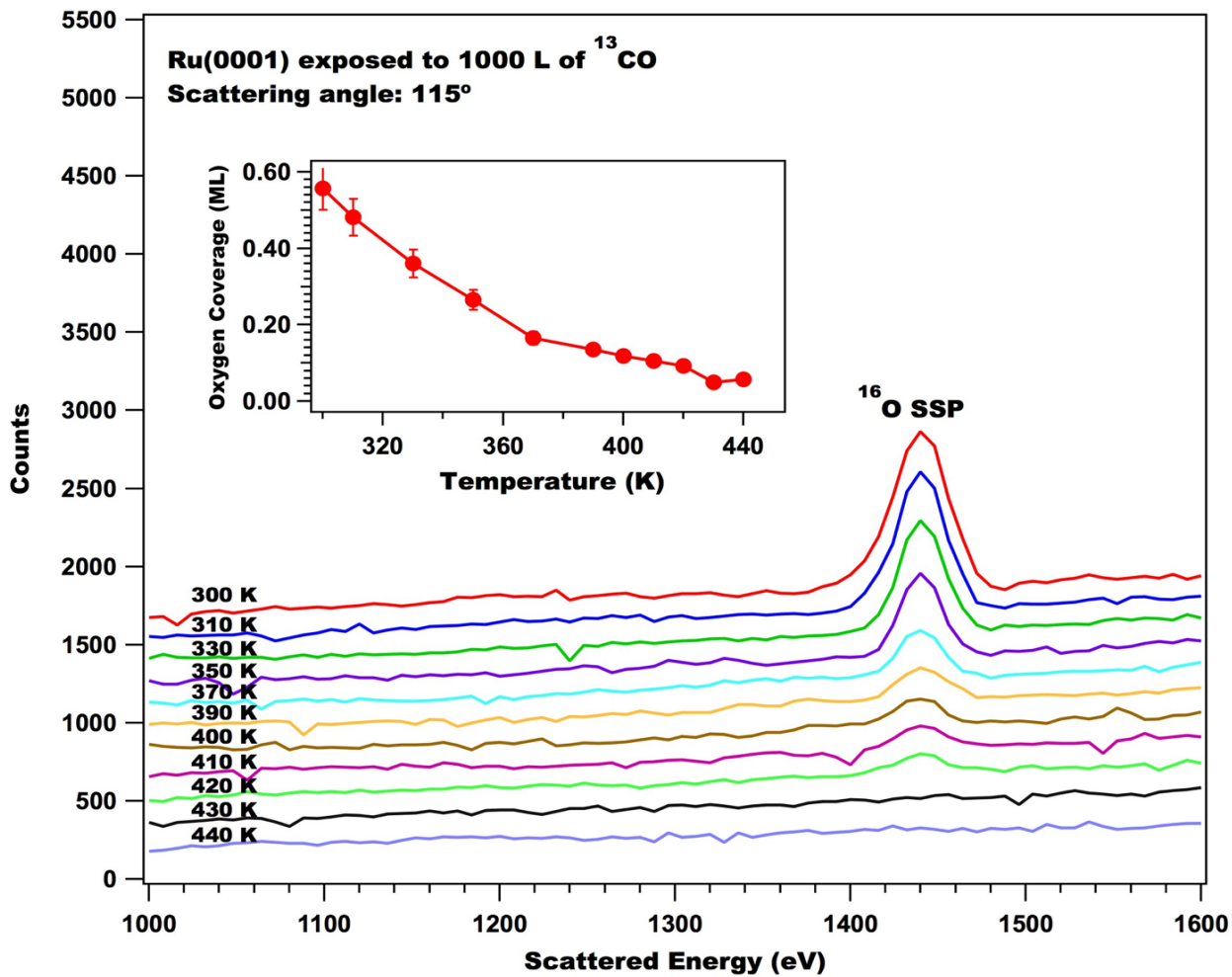

**Figure 5.** 3.0 keV He$^+$ LEIS spectra collected at a scattering angle of 115° from Ru(0001) exposed to 1000 L of $^{13}$CO at room temperature and then heated to the indicated temperatures. The uppermost spectrum was collected from as-prepared $^{13}$CO/Ru(0001). The spectra are offset from each other for clarity. The inset shows oxygen coverage as a function of the sample temperature.

The temperature dependence of $^{13}$CO adsorbed on bare Ru(0001) is also studied with LEIS, as shown in Fig. 5. The sample is exposed at room temperature to 1000 L of $^{13}$CO, followed by heating. The spectra show only the $^{16}$O SSP and no $^{13}$C peak at any time, which is consistent with the CO molecules sitting upright on the metal surface [29,31-33]. The inset shows the calculated $^{16}$O coverages versus temperature. The $^{13}$CO starts to desorb just above room temperature and the coverage decreases gradually with complete desorption occurring at 430 K, unlike the stepwise decrements observed for $^{13}$CO intercalated between Gr and Ru.

CO typically adsorbs on transition metals perpendicular to the surface plane with the carbon atom bonding to the substrate, as has been verified repeatedly [29,31-34]. This is because polarization of the molecule leads to a net negative charge on the carbon end that forms a stable chemical bond to a metal surface [35]. Thus, the observation that CO adsorbs in this same configuration when intercalated beneath Gr indicates a similar bonding to the metal.

Recent STM images and DFT calculations for CO on Ru(0001) show that at higher coverages, a fraction of the molecules adsorb at hollow sites with a slight tilt that increases with coverage up to saturation (approximately 0.6 ML) [29,36]. In the present study, 0.6 ML of $^{13}$CO is initially present on the bare Ru(0001), which should lead to some tilting. The tilting is not, however, detected with LEIS at any temperature showing that the angle is not large enough to overcome shadowing and blocking. This is in contrast to the behavior of intercalated CO, which tilts enough at elevated temperature for the $^{13}$C to be observed.

To determine the magnitude of the tilt angle required to reveal $^{13}$C, two limiting cases are considered in which the molecules tilt toward (a) the ion beam or (b) the detector. For case (a), shadowing is the dominant factor causing the $^{13}$C to be hidden. Using the Molière potential [37] and a C-O bond length of 1.13 Å, it can be derived that an angle of at least 9.8° is needed to reveal the underlying $^{13}$C atom from being shadowed by O. For case (b), the blocking of ions scattered from $^{13}$C by the O atoms would prevent them from being detected. The formula in Ref. [38] indicates that a tilt angle larger than 60° is required to make $^{13}$C completely visible to the detector. In reality, however, the tilting of the molecules occurs over the full range of azimuthal directions, so that the majority of the molecules are somewhere between the limiting (a) and (b) cases. This means that although $^{13}$C can begin to be somewhat visible at a tilt angle of 9.8°, it would not be fully visible until the angle reaches 60°.

The maximal tilt angle of CO adsorbed on bare Ru(0001) calculated from STM images, is about 15° [29]. Although this is larger than the shadowing angle calculated above, it is not large enough to sufficiently reveal the underlying C atoms in LEIS spectra collected from $^{13}$CO adsorbed on Ru(0001), which is not reasonable considering the small range of azimuthal angles that would lead to case (a) and the small scattering cross section.

For $^{13}$CO molecules intercalated between Gr and Ru(0001), however, a large change in the intensity of the $^{13}$C SSP is detected at 343 K. This shows that $^{13}$CO intercalated between Gr and Ru(0001) tilts at a larger angle than CO adsorbed on Ru, and this angle is enough to make both $^{16}$O and $^{13}$C visible. The number of visible $^{13}$C atoms is at most about 3/5 the number of visible O atoms, however, suggesting that the average tilt angle is much larger than 9.8°, but not quite 60°.

One possible cause for the tilting is that some or all of the intercalated CO lies nearly parallel to the surface on the warm samples as a precursor to dissociation, as was inferred for CO lying flat on Cr(110) [39,40]. Dissociation of CO does not appear to occur, however, as the formation of atomic O would lead to some etching of the Gr overlayer, as observed after heating Gr/Ru(0001) into which oxygen is intercalated [14,21,41,42]. No loss of C from Gr is observed (data not shown), however, and the temperature at which the CO changes its orientation is rather low. Thus, dissociation of intercalated CO is unlikely as this would infer that the C and O atoms recombine to completely recover their molecular form when cooled to room temperature since the geometry change of the intercalated molecules is reversible.

A more likely explanation is that adsorbed CO molecules covered by a graphene overlayer are less strongly bonded to the substrate at higher temperatures, and thus vibrate at larger tilt angles due to repulsion between the nearest neighboring adsorbates, as suggested in Ref. [36]. The magnitude of the tilting is also larger than for adsorption on clean Ru due to a confinement effect in which the graphene overlayer further destabilizes bonding between the intercalants and the metal substrate [12]. According to computations of the thermal vibrational modes of CO on Ru(0001) [43], both atoms vibrate parallel to the surface, but O has a larger amplitude. The lateral root mean-square displacement of O with respect to the C atom increases by roughly 0.1 Å for every 85 K increase in temperature. Because thermal vibration frequencies are much slower than the scattering times, LEIS essentially probes a frozen lattice in which the atoms are not in their equilibrium positions, but rather in a random distribution with a probability dependent on the thermal vibrational amplitudes. For scattering from vibrationally excited $^{13}$CO molecules, a larger root mean-square displacement leads to a broader distribution of O atom locations thus revealing the underlying $^{13}$C atoms.

In summary, shadowing and blocking in LEIS is used to show that CO molecules intercalate between a Gr film and a Ru(0001) substrate at room temperature with a low probability, and do not adsorb atop the Gr. The intercalated CO is upright or tilted only slightly at room temperature with the O end on top, indicating chemical bonding to Ru. The molecules tilt at a large angle at slightly elevated temperatures, however, due to increased vibrational amplitudes combined with the confining effect of the graphene overlayer. This suggests using the orientation of intercalates as a means for altering the curvature of Gr, and thus its electronic transport properties [44-46]. The reversibility of this effect could also be a means for storing information in a nano-device. In addition, since intercalated CO does not desorb until 370 K, a Gr/metal system could be used as nanocontainer to retain small molecules in vacuum.

This material is based upon work partially supported by the National Science Foundation under CHE - 1611563.

# REFERENCES

1. K. S. Novoselov, A. K. Geim, S. V. Morozov, D. Jiang, M. I. Katsnelson, I. V. Grigorieva, S. V. Dubonos, and A. A. Firsov, Nature **438**, 197 (2005).

2. C. Berger, Z. Song, X. Li, X. Wu, N. Brown, C. Naud, D. Mayou, T. Li, J. Hass, A. N. Marchenkov, E. H. Conrad, P. N. First, and W. A. de Heer, Science **312**, 1191 (2006).

3. A. K. Geim, Science **324**, 1530 (2009).

4. D. Kang, J. Y. Kwon, H. Cho, J.-H. Sim, H. S. Hwang, C. S. Kim, Y. J. Kim, R. S. Ruoff, and H. S. Shin, ACS Nano **6**, 7763 (2012).

5. J. K. Wassei and R. B. Kaner, Acc. Chem. Res. **46**, 2244 (2012).

6. H. Tetlow, J. Posthuma de Boer, I. J. Ford, D. D. Vvedensky, J. Coraux, and L. Kantorovich, Phys. Rep. **542**, 195 (2014).

7. A. T. N'Diaye, J. Coraux, T. N. Plasa, C. Busse, and T. Michely, New J. Phys. **10**, 043033 (2008).

8. P. W. Sutter, J.-I. Flege, and E. A. Sutter, Nature Materials **7**, 406 (2008).

9. Y. S. Dedkov, M. Fonin, U. Rüdiger, and C. Laubschat, Phys. Rev. Lett. **100**, 107602 (2008).

10. P. Sutter, J. T. Sadowski, and E. Sutter, Phys. Rev. B **80**, 245411 (2009).

11. D. Prasai, J. C. Tuberquia, R. R. Harl, G. K. Jennings, and K. I. Bolotin, ACS Nano **6**, 1102 (2012).

12. R. Mu, Q. Fu, L. Jin, L. Yu, G. Fang, D. Tan, and X. Bao, Angew. Chem. Int. Ed. **51**, 4856 (2012).

13. E. Grånäs, J. Knudsen, U. A. Schröder, T. Gerber, C. Busse, M. A. Arman, K. Schulte, J. N. Andersen, and T. Michely, ACS Nano **6**, 9951 (2012).

14. A. Dong, Q. Fu, M. Wei, Y. Liu, Y. Ning, F. Yang, H. Bluhm, and X. Bao, Surf. Sci. **634**, 37 (2015).

15. P. Sutter, J. T. Sadowski, and E. A. Sutter, J. Am. Chem. Soc. **132**, 8175 (2010).

16. R. Larciprete, S. Ulstrup, P. Lacovig, M. Dalmiglio, M. Bianchi, F. Mazzola, L. Hornekær, F. Orlando, A. Baraldi, P. Hofmann, and S. Lizzit, ACS Nano **6**, 9551 (2012).

17. D. Ma, Y. Zhang, M. Liu, Q. Ji, T. Gao, Y. Zhang, and Z. Liu, Nano Research **6**, 671 (2013).

18. E. Voloshina, N. Berdunov, and Y. Dedkov, Sci. Rep. **6**, 20285 (2016).

19. W. J. Rabalais, *Principles and applications of ion scattering spectrometry : surface chemical and structural analysis* (Wiley, New York, 2003).

20. H. Niehus, W. Heiland, and E. Taglauer, Surf. Sci. Rep. **17**, 213 (1993).

21. T. Li and J. A. Yarmoff, Phys. Rev. B **96**, 155441 (2017).

22. H. Zhang, Q. Fu, Y. Cui, D. Tan, and X. Bao, J. Phys. Chem. C **113**, 8296 (2009).

23. Z. Zhou, F. Gao, and D. W. Goodman, Surf. Sci. **604**, L31 (2010).

24. M. T. Robinson and I. M. Torrens, Phys. Rev. B **9**, 5008 (1974).

25. H. H. Brongersma, M. Draxler, M. de Ridder, and P. Bauer, Surf. Sci. Rep. **62**, 63 (2007).

26. S. Prusa, P. Prochazka, P. Babor, T. Sikola, R. ter Veen, M. Fartmann, T. Grehl, P. Bruner, D. Roth, P. Bauer, and H. H. Brongersma, Langmuir **31**, 9628 (2015).

27. S. N. Mikhailov, L. C. A. van den Oetelaar, and H. H. Brongersma, Nucl. Instrum. Methods Phys. Res., Sect. B **93**, 210 (1994).

28. L. C. A. van den Oetelaar, S. N. Mikhailov, and H. H. Brongersma, Nucl. Instrum. Methods Phys. Res., Sect. B **85**, 420 (1994).

29. Q. Chen, J. Liu, X. Zhou, J. Shang, Y. Zhang, X. Shao, Y. Wang, J. Li, W. Chen, G. Xu, and K. Wu, J. Phys. Chem. C **119**, 8626 (2015).

30. A. Politano and G. Chiarello, Carbon **62**, 263 (2013).

31. J. C. Fuggle, M. Steinkilberg, and D. Menzel, Chem. Phys. **11**, 307 (1975).

32. T. E. Madey, Surf. Sci. **79**, 575 (1979).

33. A. Föhlisch, M. Nyberg, J. Hasselström, O. Karis, L. G. M. Pettersson, and A. Nilsson, Phys. Rev. Lett. **85**, 3309 (2000).

34. G. Michalk, W. Moritz, H. Pfnür, and D. Menzel, Surf. Sci. **129**, 92 (1983).

35. G. E. Scuseria, M. D. Miller, F. Jensen, and J. Geertsen, J. Chem. Phys. **94**, 6660 (1991).

36. J. S. McEwen and A. Eichler, J. Chem. Phys. **126**, 094701 (2007).

37. O. S. Oen, Surface Science Letters **131**, L407 (1983).

38. A. Kutana, I. L. Bolotin, and J. W. Rabalais, Surf. Sci. **495**, 77 (2001).

39. N. D. Shinn and T. E. Madey, Phys. Rev. B **33**, 1464 (1986).

40. N. D. Shinn and T. E. Madey, J. Chem. Phys. **83**, 5928 (1985).

41. T. Li and J. A. Yarmoff, Surf. Sci. **675**, 70 (2018).

42. T. Li and J. A. Yarmoff, J. Vac. Sci. Technol. A **36**, 031404 (2018).

43. M. Gierer, H. Bludau, H. Over, and G. Ertl, Surf. Sci. **346**, 64 (1996).

44. J. C. Meyer, A. K. Geim, M. I. Katsnelson, K. S. Novoselov, T. J. Booth, and S. Roth, Nature **446**, 60 (2007).

45. A. Fasolino, J. H. Los, and M. I. Katsnelson, Nature Materials **6**, 858 (2007).

46. F. Guinea, M. I. Katsnelson, and M. A. H. Vozmediano, Phys. Rev. B **77**, 075422 (2008).